\documentclass[reprint,aps,prd]{revtex4-2}
\usepackage{amsmath}
\allowdisplaybreaks[4]
\usepackage{multirow}
\usepackage{amsfonts}
\usepackage{xcolor}
\usepackage{comment}
\usepackage{graphicx}
\usepackage{amsmath}
\usepackage{siunitx}
\usepackage{lineno}
\usepackage{float}
\usepackage{ulem}
\usepackage[
    pdfauthor={Teng Zhang},
    pdftitle={Enhancing high frequency sensitivity of gravitational wave detectors with sloshing-Sagnac interferometer},
    colorlinks=true,
    linkcolor=black,
    urlcolor=blue,
    citecolor=blue
]{hyperref}

\DeclareSIUnit{\sqrthz}{\ensuremath{\sqrt{\text{\hertz}}}}

\begin{document}
\renewcommand\texteuro{FIXME}

\allowdisplaybreaks[4]
\title{Enhancing high frequency sensitivity of gravitational wave detectors with sloshing-Sagnac interferometer}

\author{Teng Zhang}
\email{tzhang@star.sr.bham.ac.uk}
\affiliation{$^1$School of Physics and Astronomy, and Institute for Gravitational Wave Astronomy, University of Birmingham, Edgbaston, Birmingham B15\,2TT, United Kingdom}
\author{Denis Martynov}
\affiliation{$^1$School of Physics and Astronomy, and Institute for Gravitational Wave Astronomy, University of Birmingham, Edgbaston, Birmingham B15\,2TT, United Kingdom}
\author{Stefan Danilishin}
\affiliation{Department of Gravitational Waves and Fundamental Physics, Maastricht University, P.O. Box 616, 6200 MD Maastricht}
\affiliation{Nikhef, Science Park 105, 1098 XG Amsterdam, The Netherlands}
\author{Haixing Miao}
\email{haixing@star.sr.bham.ac.uk}
\affiliation{$^1$School of Physics and Astronomy, and Institute for Gravitational Wave Astronomy, University of Birmingham, Edgbaston, Birmingham B15\,2TT, United Kingdom}

\begin{abstract}
Sensitivity of gravitational-wave detectors is limited in the high-frequency band by quantum shot noise and eventually limited by the optical loss in signal recycling cavity. This limit is the main obstacle on the way to detect gravitational waves from the binary neutron star mergers in the current and the future generation detectors, as it does not depend on either the arm length, or the injected squeezing level. In this paper, we 
come up with the sloshing Sagnac interferometer topology, which can be obtained from the Michelson interferometer by optically connecting the end mirrors into an additional optical cavity. This transforms the interferometer into a triply-coupled cavity system capable of beating the loss-induced high-frequency limit of the signal-recycled Michelson interferometer.
With advanced LIGO+ comparable parameters, a sloshing-Sagnac scheme can achieve 7 times better sensitivity at 2.5\,kHz or 4 times better signal to noise ratio for a typical waveform of binary neutron post merge. Being an evolution of Michelson interferometer, the sloshing-Sagnac interferometer can possibly be used as a topology for the future generation detectors and upgrade of current detectors.

\end{abstract}
\maketitle
\section{Introduction} 
On August 17, 2017, the network of LIGO and Virgo gravitational wave (GW) detectors observed gravitational waves from a binary neutron star inspiral for the first time\,\cite{PhysRevLett.119.161101}. This observation, apart from being an unprecedented scientific breakthrough on its own right, was the first multi-messenger observation of a collision of neutron stars in both the electromagnetic and the gravitational-wave spectrum \cite{2017_ApJ.848L.12A_LVC}, which also revealed the origin of heavy chemical elements in the universe\,\cite{kasen2017origin}. Additionally, it produced a handful of exciting scientific outcomes such as,  the new test of General Relativity\cite{abbott2017gravitational}, the novel way to measure Hubble constant\,\cite{NatureHC,hotokezaka2019hubble} and a potential mechanism of the short gamma-ray bursts generation \cite{2017_ApJ.848.L13_LVC}. 

The tidal effect of the merger of binary neutron stars imprinted on gravitational waves starts to dominate from 500\,Hz onwards. The post-merger signatures are seen in the gravitational-wave signal starting from 1\,kHz.  The merger remnant may be a long-lived neutron star or can collapse into a black hole\,\cite{PhysRevD.99.102004,PhysRevD.103.044063,easter2021can}. However, the post-merger phase of gravitational-wave signal remains undetected due to insufficient sensitivity of GW detectors at higher frequencies. It is crucial however to measure this phase in order to understand the formation of the remnant star and reveal the equation of state of the nuclear matter\, \cite{PhysRevD.86.063001,PhysRevLett.113.091104,PhysRevD.91.124056}. The dominant noise of gravitational-wave detector above 1 kHz is quantum shot noise and the sensitivity is eventually limited by the vacuum fields originating from optical loss in the signal recycling cavity(SRC).

Quantum shot noise can be attributed to the vacuum fluctuations of the electromagnetic wave's amplitude and phase that stems from the fundamental uncertainty relation between energy and time, $\Delta E\Delta t\geq\frac{\hbar}{2}$. In the minimum uncertainty state, it defines the best achievable quantum noise limited sensitivity, known as quantum Cramér-Rao bound (QCRB)\,\cite{HELSTROM1967101}, which also goes by as energetic or fundamental quantum limit\,\cite{Energetic,PhysRevLett.106.090401} in the field of gravitational-wave instrument science. The  power spectral density corresponding to this limit for a Fabry–Perot--Michelson interferometer reads\,\cite{Danilishin2019,PhysRevX.9.011053}
\begin{equation}
S_{\rm QCRB}(\Omega)=\frac{c^2\hbar^2}{2S_{PP} (\Omega)L_{\rm arm}^2}\,,
\end{equation}
where $S_{PP}$ is the spectral density of power fluctuations of light in the arms, $L_{\rm arm}$ is the arm length. As follows from this formula, one can reduce the QCRB by means of increase of intracavity power or by injecting the phase-squeezed vacuum fields into the readout port of the interferometer (\textit{i.e.} amplifying the fluctuations of the amplitude quadrature).

In real-life detectors however, the QCRB is not a practical bound. The decoherence due to optical loss sets the actual limit. The loss adds extra quantum noise on top of the inherent vacuum fluctuations, and set a new bound on the sensitivity of GW detectors. By its nature, the quantum limit pertaining to loss is a limit to how much squeezing, or more generally, quantum coherence the intracavity light can possibly sustain, thereby setting a margin of usefulness of the noise cancellation schemes such as squeezing. Therefore the only way to improve the signal-to-noise ratio of the measurement apparatus beyond the loss limit is through enhancing its response to the signal. 

As demonstrated in\,\cite{PhysRevX.9.011053}, for a Fabry-Perot--Michelson interferometer, it is the optical loss in a signal recycling cavity that dominates quantum noise contribution at frequencies higher than 1 kHz due to the finite bandwidth of the interferometer response. There are several ways to improve the high-frequency signal response of the signal-recycled interferometer,\,\textit{e.g.} by taking usage of the SRC-Arm coupled cavity resonance\,\cite{PhysRevD.99.102004},  detuning the SRC \cite{PhysRevD.103.022002}, or implementing active white-light cavities\,\cite{PhysRevLett.115.211104,page2020gravitational,galaxies9010003}, or using nonlinear optical parametric amplifier inside the SRC that is known as 'quantum expander' \,\cite{korobko2019quantum}. Yet the SRC loss limit, being an internal loss of the coupled SRC-arm cavity system, is solely determined by the optical features of the arm cavity. Any quantum scheme that does not enhance the signal before it decays from arm cavity is thus marginally helpful. At frequencies beyond  the arm cavity bandwidth ($\sim$100 Hz for LIGO interferometers), the power spectral density of SRC loss reads
\begin{equation}\label{eq:SRCloss}
S_{\rm loss}^{\rm HF}(\Omega)=\frac{\hbar \Omega^2}{\omega_0P_{\rm arm}T_{\rm itm}}\epsilon_{\rm SRC}\,,
\end{equation}
where $\epsilon_{\rm SRC}$ is the round-trip loss in SRC, $\omega_0$ is the laser wavelength, $P_{\rm arm}$ is the arm cavity circulating power, $T_{\rm itm}$ is the power transmissivity of the input test mass (ITM), which determines the power ratio between the input and arms. The SRC loss-limited sensitivity at high frequencies is independent of the length of the arms. 

\begin{figure}[t]
\centering
  \includegraphics[width=1\columnwidth]{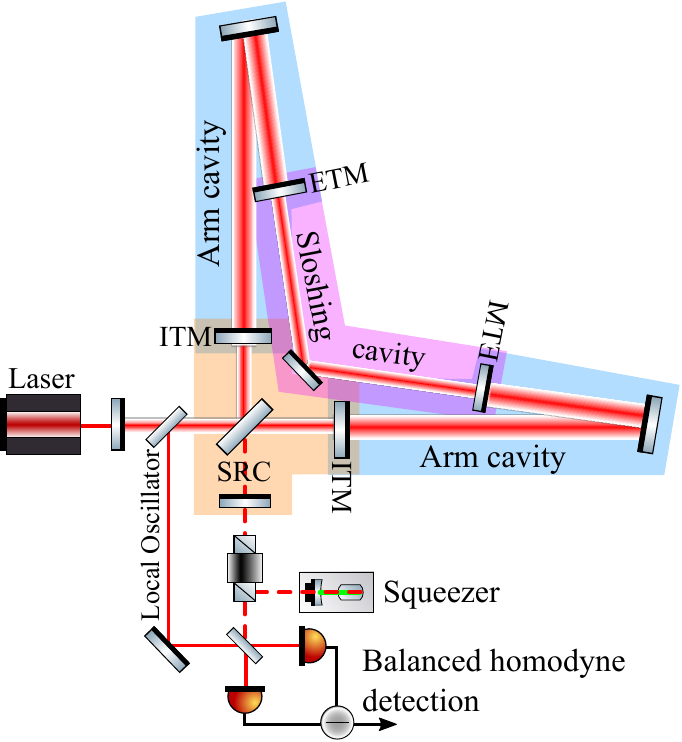}
\caption{Schematic of dual recycled sloshing-Saganac interferometer. The 6\,km arm cavity is folded within a 4\,km vacuum infrastructure. Two ETMs form a 4\,km sloshing cavity. The signal is detected with balanced homodyne readout. This high frequency configuration is shot noise limited in the whole frequency band and only requires constant phase quadrature squeezing.}
\label{fig:SloshingSagnac}
\end{figure}

Optical resonance within arms at high frequencies can be achieved by detuning the arm cavities\,\cite{Hild_2007}. However, the required detuning is so high that it leads to significant loss of circulating power and to the strong attenuation of one of the signal sidebands thereby reducing the signal response.
In this paper, we propose to use an additional sloshing cavity formed by the end test masses of the folded Fabry–Perot-Michelson interferometer to create a high-frequency resonance response. This configuration is similar to the sloshing Sagnac interferometer of \cite{Huttner_2016,Huttner_2020} with the reduced number of test masses that greatly improves the controllability of interferometer. The additional sloshing cavity can be tuned so as to enhance high frequency signal within arms and thereupon go beyond the SRC loss limited sensitivity of Fabry–Perot--Michelson interferometer. 

This paper is organised as follows. In Sec.~\ref{sec:SloshingSagnac}, we describe the characterizations of sloshing-Sagnac interferometer; in Sec.~\ref{sec:SCloss}, we describe the quantum limit from optical losses of sloshing-
Sagnac interfeormeter; in Sec.~\ref{sec:BHD}, we describe the impact of interferometer asymmetry and show the sensitivity of the interferometer.

\section{General description of Sloshing Sagnac interferometer} 
\label{sec:SloshingSagnac}
The key idea of our scheme is to enhance the high frequency signal before it decays from the arms by means of creating an additional optical resonance at high frequency. To achieve this goal, we transform the interferometer into an effective triple coupled cavity. We couple the two arms by optically connecting the end test mass (ETM) as shown in Fig.~\ref{fig:SloshingSagnac}. The new cavity formed by the two ETMs is called sloshing cavity (SC). The triple-coupled cavity system formed by the SC and the arms can be shown to have three normal modes with the central one at carrier frequency $\omega_0$ and the two modes symmetrically split around the central one by the so called sloshing frequency $\omega_s$. Using single mode approximation \cite{SD2012}, one can calculate sloshing frequency\,\cite{PhysRevD.66.122004} and bandwidth of each of the modes \cite{PhysRevD.98.044044}:
\begin{equation}\label{eq:omegasgamma}
\omega_{s}=\frac{c\sqrt{T_{\rm etm}}}{\sqrt{2L_{\rm arm}L_{c}}}\,,\gamma=\frac{cT_{\rm itm}}{4L_{\rm arm}}\,,
\end{equation}
where $T_{\rm etm}$ is the power transmissivity of the ETM, $L_{c}$ is the length of the sloshing cavity. Note that the effective bandwidth of the interferometer with signal recycling cavity is $\frac{cT_{\rm SRC}}{4L_{\rm arm}}$, where $T_{\rm SRC}$ is the effective transmissivity of the SRC formed by signal recycling mirror and ITM. The transmissivities of each arm's ITM and ETM are assumed equal to the corresponding transmissivities of the other arm's mirrors.

The propagation of light fields through the interferometer is as follows: (1) laser beams from the main beam-splitter travel through three cavities in the clockwise and counter clockwise direction and recombine at the beam-splitter akin to Sagnac interferometer; (2) the two counter-propagating waves build up power in all cavities constructively; (3) power in each arm cavity is $\frac{2}{T_{\rm itm}}$ times the laser power at the beam-splitter, which is the same as for Fabry–Perot--Michelson  interferometer; (4) power in the sloshing cavity is $\frac{(-1+\sqrt{1-T_{\rm etm}})^2(1+\sqrt{1-T_{\rm itm}})^2}{2T_{\rm itm}T_{\rm etm}}$ times that on the beam-splitter.

\begin{table}
\caption{Parameters of the interferometer}\label{ta:par}
\begin{ruledtabular}
\begin{tabular}{ccc}
    Wavelength & 1064\,nm \\
    Mirror Mass & 40\,kg \\ 
    Arm length & 6\,km (folded) \\
    SC length &4\,km\\
    Arm circulating power &800\,kW/4\,MW  \\
    Sloshing cavity power &25\,kW/125\,kW \\
    Local oscillator power/Signal port power  & 20\\
    ITM transmittivity & 0.014  \\
    ETM transmittivity & 0.12\\
    SRM transmittivity & 0.2\\
    Input squeezing level & 15\,dB \\
    Observed squeezing level (2.5\,kHz) & 6\,dB \\
    arm loss & 100\,ppm\\
    SRC loss & 1000\,ppm\\
    SC loss & 2000\,ppm\\
    Input loss & 5\%\\
    Output loss & 10\%\\ 
\end{tabular}
\end{ruledtabular}
\end{table}
\section{Loss limited sensitivity }
\label{sec:SCloss}
In single mode approximation, the power spectral density of arm loss is calculated as
\begin{equation}
S_{\rm loss}^{\rm arm}=\frac{\hbar c^2}{4\omega_0 L_{\rm arm}^2 P_{\rm arm}}\epsilon_{\rm arm}\,,
\end{equation}
where $\epsilon_{\rm arm}$ is the round trip loss in each arm cavity. The power spectral density of arm loss is the same as that of Michelson interferometer, since the arm loss mixes with signal directly in either case. Considering $100\,\rm ppm$ loss, the arm loss is not the dominating effect to the detector sensitivity.  

The power spectral density of SRC loss of the sloshing Sagnac interferometer can be calculated as
\begin{equation}
S^{\rm SRC}_{\rm loss}(\Omega)=S_{\rm shot}(\Omega)\epsilon_{\rm SRC}\,,
\end{equation}
where $\sqrt{S_{\rm shot}}$ is shot-noise-limited sensitivity of the sloshing Sagnac interferometer without SRC:
\begin{equation}
\sqrt{S_{\rm shot}(\Omega)}=\left|\frac{(\gamma\Omega -i\Omega^2+i\omega_s^2)}{\sqrt{T_{\rm itm}}\Omega}\sqrt{\frac{\hbar }{\omega_0P_{\rm arm}}}\right|\,,
\end{equation}

Compared with the noise from SRC loss of Michelson interferometer in Eq.~\ref{eq:SRCloss},  when $\Omega=\omega_s$, the SRC loss-limited sensitivity of sloshing Sagnac interferometer is improved by a factor of
\begin{equation} 
\sqrt{\frac{S_{\rm loss}^{\rm HF}(\omega_s)}{S_{\rm loss}^{\rm SRC}(\omega_s)}}=\frac{\omega_{s}}{\gamma}\,.
\end{equation}
Here $\omega_s$, which is around 2.5\,kHz$\times 2\pi$ in this paper is much larger than the arm cavity bandwidth $\gamma$, which is around 40\,Hz$\times 2\pi$ for advanced LIGO. 

The sensitivity of the sloshing Sagnac interferometer is actually limited by the new loss from SC and its power spectral density is calculated as
\begin{equation}
S_{\rm loss}^{\rm SC}=\frac{\hbar \omega_s^4 }{2\omega_0 P_{\rm arm}T_{\rm etm}\Omega^2}\epsilon_{\rm SC}\,,
\end{equation}
where $\epsilon_{\rm SC}$ is the round trip loss in SC. 
Compared with Eq.~\ref{eq:SRCloss}, at $\Omega=\omega_s$, the loss limited sensitivity of new scheme can be improved by a factor of
\begin{equation}
\sqrt{\frac{S_{\rm loss}^{\rm HF}(\omega_s)}{S_{\rm loss}^{\rm SC}(\omega_s)}}=\sqrt{\frac{T_{\rm etm}\epsilon_{\rm SC}}{2T_{\rm itm}\epsilon_{\rm SRC}}}\,.
\end{equation}
Note that the sensitivity limit stemming from losses in either SRC, or SC is independent of the arm length in single mode approximation.

\begin{figure}[t]
\centering
  \includegraphics[width=1\columnwidth]{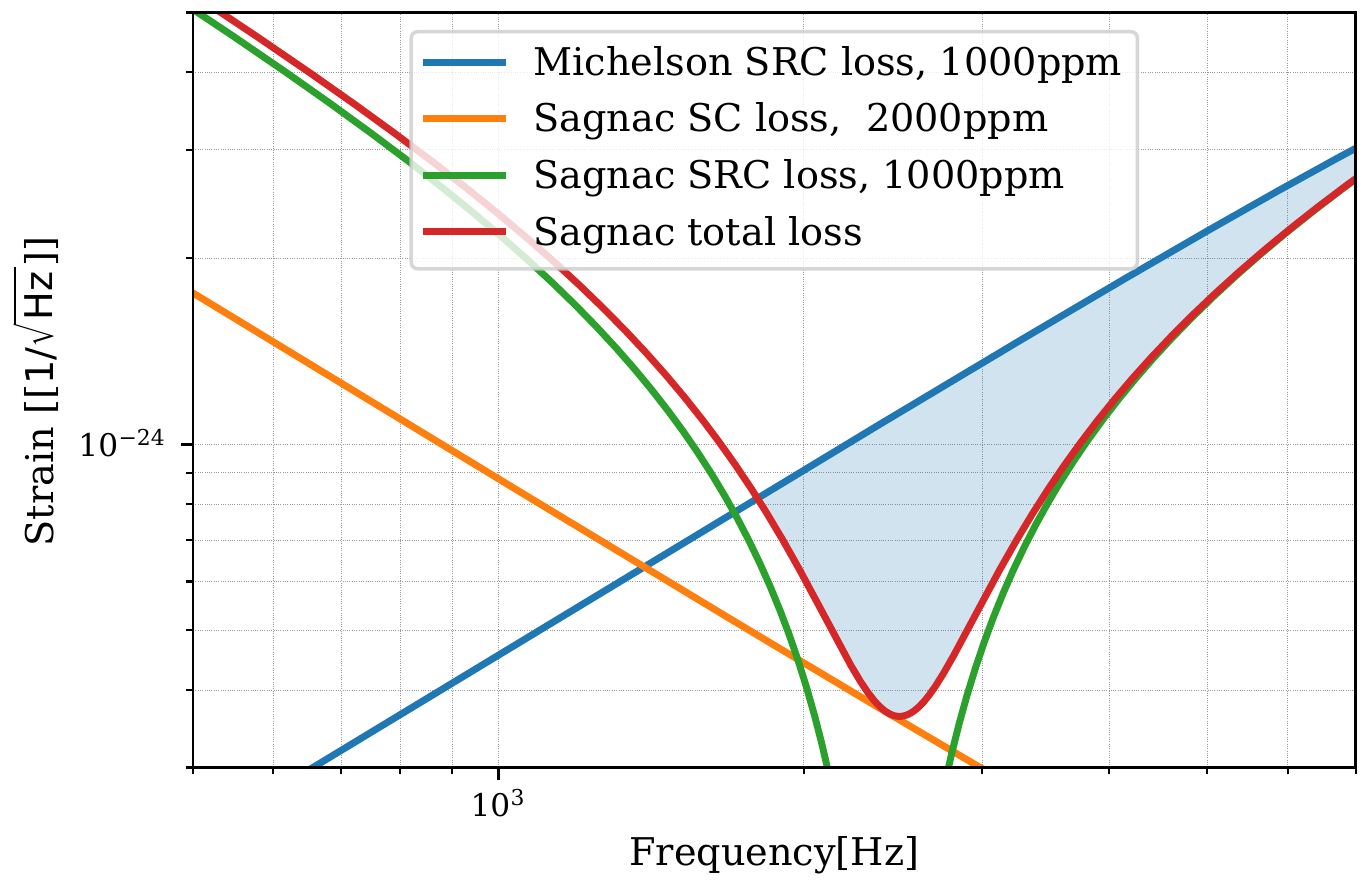}
\caption{Loss limited sensitivity of  sloshing-Sagnac interferometer with arm power 800\.kW.
The red line represents the sensitivity limit from total loss, which is formed by the SRC loss (red) and SC loss (green). The blue shading area denotes the improved sensitivity limit from Michelson interferometer to Sloshing-Sagnac interferometer. The parameters used are listed in Table~\ref{ta:par}.
}
\label{fig:loss}
\end{figure}

As described in\,\cite{PhysRevD.99.102004}, the main contribution to the SRC loss comes from the wavefront distortion at the ITMs due to thermal lensing effect. The effective SRC loss in the differential mode is only half of the total loss on both ITMs. The effective SC loss however is defined as the total loss from both ETMs. If we assume that $T_{\rm itm}=T_{\rm etm}$ and  $\epsilon_{\rm SC}=2\epsilon_{\rm SRC}$, then the SC loss-limited sensitivity of sloshing Sagnac is equal to that of the signal-recycled Fabry-Perot--Michelson interferometer at $\omega_s$. One can conclude also that better sensitivity can be achieved at higher values of $T_{\rm etm}$. Taking the Advanced LIGO parameters, we can show the improvement of loss limited sensitivity provided by the sloshing Sagnac scheme as shaded area in Fig.~\ref{fig:loss}. Here $T_{\rm etm}$ is chosen to be equal to 0.12. The circulating power in SC is 4.5 times of the laser power at the beam-splitter. Other parameters are shown in Table~\ref{ta:par}.

\begin{figure}[t]
\centering
  \includegraphics[width=1\columnwidth]{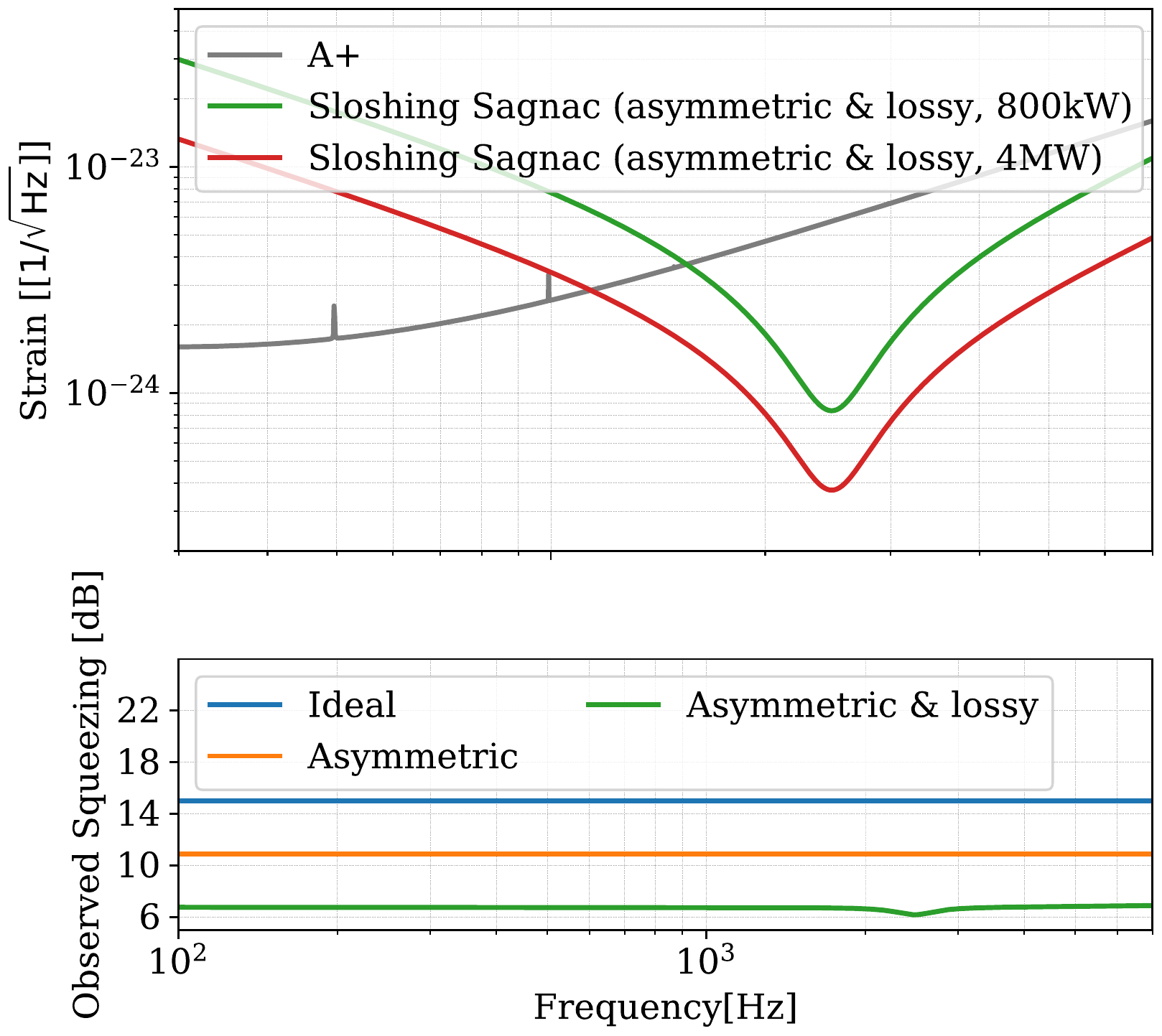}
\caption{Upper panel: Sensitivity of sloshing-Sagnac interferometer with different circulating power in the arms. The grey line is the A+ design sensitivity. The green and the red lines are the corresponding sensitivities of the sloshing-Saganc interferometer with the arm power 800\,kW and 4\,MW, respectively. The parameters are listed in Table~\ref{ta:par}. Lower panel: The observed squeezing of 800\,kW interferometer in different cases with 15\,dB injection squeezing. The blue line is the ideal case, the orange line indicates the impact caused by asymmetries when the ratio of local oscillator power over beam power at signal port is 20. The green line includes both asymmetries and losses. }
\label{fig:HF}
\end{figure}
\section{Interferometer Asymmetry} \label{sec:BHD}
Sagnac interferometer is highly sensitive to the symmetry of the arms and of the beam-splitter \cite{2015asymSag}, however the influence of such asymmetries as the skewed beam-splitter ratio and impedance mismatch between the ITMs/ETMs in two arms is mainly limited to radiation pressure noise dominated sensitivity. In the shot noise limited high frequency detector and the frequency band of interest ($\textgreater$ 1\,kHz),  the influence on detector response is negligible as long as the arm power is kept at the design level. 

In our design, we implement balanced homodyne readout.  
For phase quadrature measurement, the photocurrent of general balanced homodyne readout can be written as\,\cite{Zhang_2018}
\begin{equation}\label{eq:I}
I=L \hat{o}_{s}-O \hat{l}_{s}\,,
\end{equation}
where $L$ is the local oscillator amplitude, $O$ is the light amplitude at the signal port, $\hat{o}_s$ is the phase fluctuation at the signal port including both noise and gravitational-wave signal, $\hat{l}_s$ is the local oscillator phase noise. Due to asymmetry, the laser stationary component and noise fluctuations will couple to the interferometer  signal port. At the same time, the squeezed vacuum from the dark port gets lost to the laser port. Calculation shows however that coupling factor between the laser port and the readout port is so tiny that the corresponding loss effect can be ignored. For example, the coupling from the laser port to the readout port due to beam-splitter imbalance is $|R_{\rm BS}-T_{\rm BS}|^2$, where $R_{\rm BS}$ and $T_{\rm BS}$ are the beam-splitter power reflectivity and transmissivity, respectively. However, the absolute power that appears at signal port, after signal recycling mirror is important. In addition, note that the signal recycling cavity will further amplify the power of light that couples through the beam-splitter by $\sim 4/{T_{\rm SRM}}$, where $T_{\rm SRM}$ is the signal recycling mirror transmissivity.  To maintain a good signal to noise ratio, the second term in Eq.~\ref{eq:I} needs to be much smaller than the first term, \textit{i.e.} $L\gg O$ as usual. With only asymmetry effect taken into account, the observed noise level is $\frac{|O|^2}{|L|^2}+e^{-2r}$, where $e^{-2r}=0.1$ means 10\,dB squeezing.
We simulate the leaking power to dark port of the interferometer including optical losses with simulation software {FINESSE}\,\cite{Freise_2004}. Assuming $0.1\%$ asymmetry between two ITMs/ETMs and between the transmissivity and reflectivivity of beam-splitter, the resulting leaking power at signal port is $\sim$ 20\,mW. The asymmetry is definded as the differential part over the average of two items. In this case, implementing 400\,mW local oscillator, with only asymmetry effect taken into account, the observed noise level under 15\,dB injection squeezing is expected to be $10.8\,$dB.
For $1\%$ asymmetry, the leaking power at signal power will be $\sim 2$\,W. It could lead large number of photodetectors to bear the required even much higher power from local oscillator. One solution is to split the high frequency signal sidebands and DC light with a so called amplitude filter cavity\,\cite{PhysRevD.102.102003} at the interferometer output. The amplitude filter cavity is a critically coupled cavity on resonance at DC and with bandwidth much smaller than $\omega_s$. Hereby, the laser at DC will transmit through and the to be detected high frequency sidebands are reflected from the filter cavity. This additional filter scheme could almost eliminate asymmetry impact. In this work, we assumes a factor of 20 higher local oscillator power than the beam power at signal port which is also what $0.1\%$ asymmetry without amplitude filter cavity can achieve.
The orange line in the lower panel of Fig.~\ref{fig:HF} shows $\sim$10.8\,dB observed squeezing considering asymmetries. Including losses, the observed squeezing is shown as the green line in the lower panel of Fig.~\ref{fig:HF}. The results are carried out through FINESSE with parameters listed in Table~\ref{ta:par}. The sensitivity of the detector is shot noise limited in the whole frequency band, and the peak sensitivity is a factor 7 better than A+ design sensitivity\,\cite{T1800042} at $\omega_s\approx2.5\,\rm kHz\times2\pi$ with effective bandwidth 445\,Hz. Taking a typical waveform of peak mode of binary neutron star post merge in\,\cite{PhysRevD.103.044063}, the resulting signal to noise ratio is a factor 4 better than that from A+.
As a compaision to the LIGO-HF design in\,\cite{PhysRevD.98.044044}, an example of 4\,MW arm power is also shown in Fig.~\ref{fig:HF}.

\section{Conclusion and discussion}
In this paper, we introduce a new type of interferometer, sloshing-Sagnac interferometer, which is able to improve the kilo-Hertz sensitivity beyond the sensitivity limit from SRC losses of Michelson interferometer. This relies on the enhanced signal response at resonant mode of triply-coupled-cavities, which mitigates the contamination from losses outside of the coupled cavities. Meanwhile, the new induced internal loss, SC loss, can be mitigated through lowering the finesse of the sloshing cavity. 
In the interferometer, the macroscopic cavity lengths need to be determined to fulfil the required resonant frequency. Arm cavities with 6\,km length and sloshing cavity with 4\,km length can possibly be folded into a 4km facility the same as LIGO facility. Assuming the same beam size as that of a 4\,km configuration with straight arms, the folded configuration gives similar mirror thermal noise\,\cite{Sanders_2016}.

As an outlook, the sloshing-Sagnac studied in this paper is also a speed meter.
A speed meter type low frequency response following $\Omega$ can be observed in Fig.~\ref{fig:HF}.  Exploiting a different part of the parameter space, \textit{i.e.} lower $\omega_s$, the low frequency signal will be enhanced and the sensitivity will be limited by radiation pressure noise. Compared with Michelson interferometer, speed meter has lower radiation pressure noise. Hereby, this type of sloshing-Sagnac can also serve as a broadband low frequency detector.

\section{Acknowledgements} 
We thank the support from simulation software, FINESSE. We thank Yanbei Chen, Mikhail. Korobko for fruitful discussion. 
T. Z., D. M. and H. M. acknowledge the support of the Institute for Gravitational Wave Astronomy at the University of Birmingham, STFC Quantum Technology for Fundamental Physics scheme (Grant No. ST/T006609/1), and EPSRC New Horizon Scheme (Grant No. EP/V048872/1
)  H. M. is supported by UK STFC Ernest Rutherford Fellowship (Grant No. ST/M005844/11). S.D. acknowledges the support of the Faculty of Science and Engineering of the Maastricht University.


\bibliography{bibliography}

\begin{thebibliography}{35}%
\makeatletter
\providecommand \@ifxundefined [1]{%
 \@ifx{#1\undefined}
}%
\providecommand \@ifnum [1]{%
 \ifnum #1\expandafter \@firstoftwo
 \else \expandafter \@secondoftwo
 \fi
}%
\providecommand \@ifx [1]{%
 \ifx #1\expandafter \@firstoftwo
 \else \expandafter \@secondoftwo
 \fi
}%
\providecommand \natexlab [1]{#1}%
\providecommand \enquote  [1]{``#1''}%
\providecommand \bibnamefont  [1]{#1}%
\providecommand \bibfnamefont [1]{#1}%
\providecommand \citenamefont [1]{#1}%
\providecommand \href@noop [0]{\@secondoftwo}%
\providecommand \href [0]{\begingroup \@sanitize@url \@href}%
\providecommand \@href[1]{\@@startlink{#1}\@@href}%
\providecommand \@@href[1]{\endgroup#1\@@endlink}%
\providecommand \@sanitize@url [0]{\catcode `\\12\catcode `\$12\catcode
  `\&12\catcode `\#12\catcode `\^12\catcode `\_12\catcode `\%12\relax}%
\providecommand \@@startlink[1]{}%
\providecommand \@@endlink[0]{}%
\providecommand \url  [0]{\begingroup\@sanitize@url \@url }%
\providecommand \@url [1]{\endgroup\@href {#1}{\urlprefix }}%
\providecommand \urlprefix  [0]{URL }%
\providecommand \Eprint [0]{\href }%
\providecommand \doibase [0]{https://doi.org/}%
\providecommand \selectlanguage [0]{\@gobble}%
\providecommand \bibinfo  [0]{\@secondoftwo}%
\providecommand \bibfield  [0]{\@secondoftwo}%
\providecommand \translation [1]{[#1]}%
\providecommand \BibitemOpen [0]{}%
\providecommand \bibitemStop [0]{}%
\providecommand \bibitemNoStop [0]{.\EOS\space}%
\providecommand \EOS [0]{\spacefactor3000\relax}%
\providecommand \BibitemShut  [1]{\csname bibitem#1\endcsname}%
\let\auto@bib@innerbib\@empty
\bibitem [{\citenamefont {Abbott}\ \emph {et~al.}(2017)\citenamefont {Abbott},
  \citenamefont {Abbott}, \citenamefont {Abbott} \emph
  {et~al.}}]{PhysRevLett.119.161101}%
  \BibitemOpen
  \bibfield  {author} {\bibinfo {author} {\bibfnamefont {B.~P.}\ \bibnamefont
  {Abbott}}, \bibinfo {author} {\bibfnamefont {R.}~\bibnamefont {Abbott}},
  \bibinfo {author} {\bibfnamefont {T.~D.}\ \bibnamefont {Abbott}}, \emph
  {et~al.} (\bibinfo {collaboration} {LIGO Scientific Collaboration and Virgo
  Collaboration}),\ }\bibfield  {title} {\bibinfo {title} {Gw170817:
  Observation of gravitational waves from a binary neutron star inspiral},\
  }\href {https://doi.org/10.1103/PhysRevLett.119.161101} {\bibfield  {journal}
  {\bibinfo  {journal} {Phys. Rev. Lett.}\ }\textbf {\bibinfo {volume} {119}},\
  \bibinfo {pages} {161101} (\bibinfo {year} {2017})}\BibitemShut {NoStop}%
\bibitem [{\citenamefont {{Abbott}}\ \emph {et~al.}(2017)\citenamefont
  {{Abbott}}, \citenamefont {{Abbott}}, \citenamefont {{Abbott}}, \citenamefont
  {{Acernese}}, \citenamefont {{Ackley}}, \citenamefont {{Adams}},
  \citenamefont {{Adams}}, \citenamefont {{Addesso}}, \citenamefont
  {{Adhikari}}, \citenamefont {{Adya}},\ and\ \citenamefont
  {et~al.}}]{2017_ApJ.848L.12A_LVC}%
  \BibitemOpen
  \bibfield  {author} {\bibinfo {author} {\bibfnamefont {B.~P.}\ \bibnamefont
  {{Abbott}}}, \bibinfo {author} {\bibfnamefont {R.}~\bibnamefont {{Abbott}}},
  \bibinfo {author} {\bibfnamefont {T.~D.}\ \bibnamefont {{Abbott}}}, \bibinfo
  {author} {\bibfnamefont {F.}~\bibnamefont {{Acernese}}}, \bibinfo {author}
  {\bibfnamefont {K.}~\bibnamefont {{Ackley}}}, \bibinfo {author}
  {\bibfnamefont {C.}~\bibnamefont {{Adams}}}, \bibinfo {author} {\bibfnamefont
  {T.}~\bibnamefont {{Adams}}}, \bibinfo {author} {\bibfnamefont
  {P.}~\bibnamefont {{Addesso}}}, \bibinfo {author} {\bibfnamefont {R.~X.}\
  \bibnamefont {{Adhikari}}}, \bibinfo {author} {\bibfnamefont {V.~B.}\
  \bibnamefont {{Adya}}},\ and\ \bibinfo {author} {\bibnamefont {et~al.}},\
  }\bibfield  {title} {\bibinfo {title} {{Multi-messenger Observations of a
  Binary Neutron Star Merger}},\ }\href
  {https://doi.org/10.3847/2041-8213/aa91c9} {\bibfield  {journal} {\bibinfo
  {journal} {Astrophys. J. Lett.}\ }\textbf {\bibinfo {volume} {848}},\
  \bibinfo {eid} {L12} (\bibinfo {year} {2017})},\ \Eprint
  {https://arxiv.org/abs/1710.05833} {arXiv:1710.05833 [astro-ph.HE]}
  \BibitemShut {NoStop}%
\bibitem [{\citenamefont {Kasen}\ \emph {et~al.}(2017)\citenamefont {Kasen},
  \citenamefont {Metzger}, \citenamefont {Barnes}, \citenamefont {Quataert},\
  and\ \citenamefont {Ramirez-Ruiz}}]{kasen2017origin}%
  \BibitemOpen
  \bibfield  {author} {\bibinfo {author} {\bibfnamefont {D.}~\bibnamefont
  {Kasen}}, \bibinfo {author} {\bibfnamefont {B.}~\bibnamefont {Metzger}},
  \bibinfo {author} {\bibfnamefont {J.}~\bibnamefont {Barnes}}, \bibinfo
  {author} {\bibfnamefont {E.}~\bibnamefont {Quataert}},\ and\ \bibinfo
  {author} {\bibfnamefont {E.}~\bibnamefont {Ramirez-Ruiz}},\ }\bibfield
  {title} {\bibinfo {title} {Origin of the heavy elements in binary
  neutron-star mergers from a gravitational-wave event},\ }\href@noop {}
  {\bibfield  {journal} {\bibinfo  {journal} {Nature}\ }\textbf {\bibinfo
  {volume} {551}},\ \bibinfo {pages} {80} (\bibinfo {year} {2017})}\BibitemShut
  {NoStop}%
\bibitem [{\citenamefont {Abbott}\ \emph {et~al.}(2017)\citenamefont {Abbott},
  \citenamefont {Abbott}, \citenamefont {Abbott}, \citenamefont {Acernese},
  \citenamefont {Ackley}, \citenamefont {Adams}, \citenamefont {Adams},
  \citenamefont {Addesso}, \citenamefont {Adhikari}, \citenamefont {Adya} \emph
  {et~al.}}]{abbott2017gravitational}%
  \BibitemOpen
  \bibfield  {author} {\bibinfo {author} {\bibfnamefont {B.~P.}\ \bibnamefont
  {Abbott}}, \bibinfo {author} {\bibfnamefont {R.}~\bibnamefont {Abbott}},
  \bibinfo {author} {\bibfnamefont {T.}~\bibnamefont {Abbott}}, \bibinfo
  {author} {\bibfnamefont {F.}~\bibnamefont {Acernese}}, \bibinfo {author}
  {\bibfnamefont {K.}~\bibnamefont {Ackley}}, \bibinfo {author} {\bibfnamefont
  {C.}~\bibnamefont {Adams}}, \bibinfo {author} {\bibfnamefont
  {T.}~\bibnamefont {Adams}}, \bibinfo {author} {\bibfnamefont
  {P.}~\bibnamefont {Addesso}}, \bibinfo {author} {\bibfnamefont
  {R.}~\bibnamefont {Adhikari}}, \bibinfo {author} {\bibfnamefont
  {V.}~\bibnamefont {Adya}}, \emph {et~al.},\ }\bibfield  {title} {\bibinfo
  {title} {Gravitational waves and gamma-rays from a binary neutron star
  merger: Gw170817 and grb 170817a},\ }\href@noop {} {\bibfield  {journal}
  {\bibinfo  {journal} {The Astrophysical Journal Letters}\ }\textbf {\bibinfo
  {volume} {848}},\ \bibinfo {pages} {L13} (\bibinfo {year}
  {2017})}\BibitemShut {NoStop}%
\bibitem [{\citenamefont {Collaboration}\ and\ \citenamefont
  {Collaboration}(2017)}]{NatureHC}%
  \BibitemOpen
  \bibfield  {author} {\bibinfo {author} {\bibfnamefont {T.~L.~S.}\
  \bibnamefont {Collaboration}}\ and\ \bibinfo {author} {\bibfnamefont {T.~V.}\
  \bibnamefont {Collaboration}},\ }\bibfield  {title} {\bibinfo {title} {A
  gravitational-wave standard siren measurement of the hubble constant},\
  }\href {https://doi.org/10.1038/nature24471} {\bibfield  {journal} {\bibinfo
  {journal} {Nature}\ }\textbf {\bibinfo {volume} {551}},\ \bibinfo {pages}
  {85} (\bibinfo {year} {2017})}\BibitemShut {NoStop}%
\bibitem [{\citenamefont {Hotokezaka}\ \emph {et~al.}(2019)\citenamefont
  {Hotokezaka}, \citenamefont {Nakar}, \citenamefont {Gottlieb}, \citenamefont
  {Nissanke}, \citenamefont {Masuda}, \citenamefont {Hallinan}, \citenamefont
  {Mooley},\ and\ \citenamefont {Deller}}]{hotokezaka2019hubble}%
  \BibitemOpen
  \bibfield  {author} {\bibinfo {author} {\bibfnamefont {K.}~\bibnamefont
  {Hotokezaka}}, \bibinfo {author} {\bibfnamefont {E.}~\bibnamefont {Nakar}},
  \bibinfo {author} {\bibfnamefont {O.}~\bibnamefont {Gottlieb}}, \bibinfo
  {author} {\bibfnamefont {S.}~\bibnamefont {Nissanke}}, \bibinfo {author}
  {\bibfnamefont {K.}~\bibnamefont {Masuda}}, \bibinfo {author} {\bibfnamefont
  {G.}~\bibnamefont {Hallinan}}, \bibinfo {author} {\bibfnamefont {K.~P.}\
  \bibnamefont {Mooley}},\ and\ \bibinfo {author} {\bibfnamefont {A.~T.}\
  \bibnamefont {Deller}},\ }\bibfield  {title} {\bibinfo {title} {A hubble
  constant measurement from superluminal motion of the jet in gw170817},\
  }\href@noop {} {\bibfield  {journal} {\bibinfo  {journal} {Nature Astronomy}\
  }\textbf {\bibinfo {volume} {3}},\ \bibinfo {pages} {940} (\bibinfo {year}
  {2019})}\BibitemShut {NoStop}%
\bibitem [{\citenamefont {{Abbott}}\ \emph {et~al.}(2017)\citenamefont
  {{Abbott}}, \citenamefont {{Abbott}}, \citenamefont {{Abbott}}, \citenamefont
  {{Acernese}}, \citenamefont {{Ackley}}, \citenamefont {{Adams}},
  \citenamefont {{Adams}}, \citenamefont {{Addesso}}, \citenamefont
  {{Adhikari}}, \citenamefont {{Adya}},\ and\ \citenamefont
  {et~al.}}]{2017_ApJ.848.L13_LVC}%
  \BibitemOpen
  \bibfield  {author} {\bibinfo {author} {\bibfnamefont {B.~P.}\ \bibnamefont
  {{Abbott}}}, \bibinfo {author} {\bibfnamefont {R.}~\bibnamefont {{Abbott}}},
  \bibinfo {author} {\bibfnamefont {T.~D.}\ \bibnamefont {{Abbott}}}, \bibinfo
  {author} {\bibfnamefont {F.}~\bibnamefont {{Acernese}}}, \bibinfo {author}
  {\bibfnamefont {K.}~\bibnamefont {{Ackley}}}, \bibinfo {author}
  {\bibfnamefont {C.}~\bibnamefont {{Adams}}}, \bibinfo {author} {\bibfnamefont
  {T.}~\bibnamefont {{Adams}}}, \bibinfo {author} {\bibfnamefont
  {P.}~\bibnamefont {{Addesso}}}, \bibinfo {author} {\bibfnamefont {R.~X.}\
  \bibnamefont {{Adhikari}}}, \bibinfo {author} {\bibfnamefont {V.~B.}\
  \bibnamefont {{Adya}}},\ and\ \bibinfo {author} {\bibnamefont {et~al.}},\
  }\bibfield  {title} {\bibinfo {title} {{Gravitational Waves and Gamma-Rays
  from a Binary Neutron Star Merger: GW170817 and GRB 170817A}},\ }\href
  {https://doi.org/10.3847/2041-8213/aa920c} {\bibfield  {journal} {\bibinfo
  {journal} {Astrophys. J. Lett.}\ }\textbf {\bibinfo {volume} {848}},\
  \bibinfo {eid} {L13} (\bibinfo {year} {2017})},\ \Eprint
  {https://arxiv.org/abs/1710.05834} {arXiv:1710.05834 [astro-ph.HE]}
  \BibitemShut {NoStop}%
\bibitem [{\citenamefont {Martynov}\ \emph {et~al.}(2019)\citenamefont
  {Martynov}, \citenamefont {Miao}, \citenamefont {Yang}, \citenamefont
  {Vivanco}, \citenamefont {Thrane}, \citenamefont {Smith}, \citenamefont
  {Lasky}, \citenamefont {East}, \citenamefont {Adhikari}, \citenamefont
  {Bauswein}, \citenamefont {Brooks}, \citenamefont {Chen}, \citenamefont
  {Corbitt}, \citenamefont {Freise}, \citenamefont {Grote}, \citenamefont
  {Levin}, \citenamefont {Zhao},\ and\ \citenamefont
  {Vecchio}}]{PhysRevD.99.102004}%
  \BibitemOpen
  \bibfield  {author} {\bibinfo {author} {\bibfnamefont {D.}~\bibnamefont
  {Martynov}}, \bibinfo {author} {\bibfnamefont {H.}~\bibnamefont {Miao}},
  \bibinfo {author} {\bibfnamefont {H.}~\bibnamefont {Yang}}, \bibinfo {author}
  {\bibfnamefont {F.~H.}\ \bibnamefont {Vivanco}}, \bibinfo {author}
  {\bibfnamefont {E.}~\bibnamefont {Thrane}}, \bibinfo {author} {\bibfnamefont
  {R.}~\bibnamefont {Smith}}, \bibinfo {author} {\bibfnamefont
  {P.}~\bibnamefont {Lasky}}, \bibinfo {author} {\bibfnamefont {W.~E.}\
  \bibnamefont {East}}, \bibinfo {author} {\bibfnamefont {R.}~\bibnamefont
  {Adhikari}}, \bibinfo {author} {\bibfnamefont {A.}~\bibnamefont {Bauswein}},
  \bibinfo {author} {\bibfnamefont {A.}~\bibnamefont {Brooks}}, \bibinfo
  {author} {\bibfnamefont {Y.}~\bibnamefont {Chen}}, \bibinfo {author}
  {\bibfnamefont {T.}~\bibnamefont {Corbitt}}, \bibinfo {author} {\bibfnamefont
  {A.}~\bibnamefont {Freise}}, \bibinfo {author} {\bibfnamefont
  {H.}~\bibnamefont {Grote}}, \bibinfo {author} {\bibfnamefont
  {Y.}~\bibnamefont {Levin}}, \bibinfo {author} {\bibfnamefont
  {C.}~\bibnamefont {Zhao}},\ and\ \bibinfo {author} {\bibfnamefont
  {A.}~\bibnamefont {Vecchio}},\ }\bibfield  {title} {\bibinfo {title}
  {Exploring the sensitivity of gravitational wave detectors to neutron star
  physics},\ }\href {https://doi.org/10.1103/PhysRevD.99.102004} {\bibfield
  {journal} {\bibinfo  {journal} {Phys. Rev. D}\ }\textbf {\bibinfo {volume}
  {99}},\ \bibinfo {pages} {102004} (\bibinfo {year} {2019})}\BibitemShut
  {NoStop}%
\bibitem [{\citenamefont {Zhang}\ \emph
  {et~al.}(2021{\natexlab{a}})\citenamefont {Zhang}, \citenamefont {Smetana},
  \citenamefont {Chen}, \citenamefont {Bentley}, \citenamefont {Martynov},
  \citenamefont {Miao}, \citenamefont {East},\ and\ \citenamefont
  {Yang}}]{PhysRevD.103.044063}%
  \BibitemOpen
  \bibfield  {author} {\bibinfo {author} {\bibfnamefont {T.}~\bibnamefont
  {Zhang}}, \bibinfo {author} {\bibfnamefont {J.~c.~v.}\ \bibnamefont
  {Smetana}}, \bibinfo {author} {\bibfnamefont {Y.}~\bibnamefont {Chen}},
  \bibinfo {author} {\bibfnamefont {J.}~\bibnamefont {Bentley}}, \bibinfo
  {author} {\bibfnamefont {D.}~\bibnamefont {Martynov}}, \bibinfo {author}
  {\bibfnamefont {H.}~\bibnamefont {Miao}}, \bibinfo {author} {\bibfnamefont
  {W.~E.}\ \bibnamefont {East}},\ and\ \bibinfo {author} {\bibfnamefont
  {H.}~\bibnamefont {Yang}},\ }\bibfield  {title} {\bibinfo {title} {Toward
  observing neutron star collapse with gravitational wave detectors},\ }\href
  {https://doi.org/10.1103/PhysRevD.103.044063} {\bibfield  {journal} {\bibinfo
   {journal} {Phys. Rev. D}\ }\textbf {\bibinfo {volume} {103}},\ \bibinfo
  {pages} {044063} (\bibinfo {year} {2021}{\natexlab{a}})}\BibitemShut
  {NoStop}%
\bibitem [{\citenamefont {Easter}\ \emph {et~al.}(2021)\citenamefont {Easter},
  \citenamefont {Lasky},\ and\ \citenamefont {Casey}}]{easter2021can}%
  \BibitemOpen
  \bibfield  {author} {\bibinfo {author} {\bibfnamefont {P.~J.}\ \bibnamefont
  {Easter}}, \bibinfo {author} {\bibfnamefont {P.~D.}\ \bibnamefont {Lasky}},\
  and\ \bibinfo {author} {\bibfnamefont {A.~R.}\ \bibnamefont {Casey}},\
  }\bibfield  {title} {\bibinfo {title} {Can we measure the collapse time of a
  post-merger remnant for a future gw170817-like event?},\ }\href@noop {}
  {\bibfield  {journal} {\bibinfo  {journal} {arXiv preprint arXiv:2106.04064}\
  } (\bibinfo {year} {2021})}\BibitemShut {NoStop}%
\bibitem [{\citenamefont {Bauswein}\ \emph {et~al.}(2012)\citenamefont
  {Bauswein}, \citenamefont {Janka}, \citenamefont {Hebeler},\ and\
  \citenamefont {Schwenk}}]{PhysRevD.86.063001}%
  \BibitemOpen
  \bibfield  {author} {\bibinfo {author} {\bibfnamefont {A.}~\bibnamefont
  {Bauswein}}, \bibinfo {author} {\bibfnamefont {H.-T.}\ \bibnamefont {Janka}},
  \bibinfo {author} {\bibfnamefont {K.}~\bibnamefont {Hebeler}},\ and\ \bibinfo
  {author} {\bibfnamefont {A.}~\bibnamefont {Schwenk}},\ }\bibfield  {title}
  {\bibinfo {title} {Equation-of-state dependence of the gravitational-wave
  signal from the ring-down phase of neutron-star mergers},\ }\href
  {https://doi.org/10.1103/PhysRevD.86.063001} {\bibfield  {journal} {\bibinfo
  {journal} {Phys. Rev. D}\ }\textbf {\bibinfo {volume} {86}},\ \bibinfo
  {pages} {063001} (\bibinfo {year} {2012})}\BibitemShut {NoStop}%
\bibitem [{\citenamefont {Takami}\ \emph {et~al.}(2014)\citenamefont {Takami},
  \citenamefont {Rezzolla},\ and\ \citenamefont
  {Baiotti}}]{PhysRevLett.113.091104}%
  \BibitemOpen
  \bibfield  {author} {\bibinfo {author} {\bibfnamefont {K.}~\bibnamefont
  {Takami}}, \bibinfo {author} {\bibfnamefont {L.}~\bibnamefont {Rezzolla}},\
  and\ \bibinfo {author} {\bibfnamefont {L.}~\bibnamefont {Baiotti}},\
  }\bibfield  {title} {\bibinfo {title} {Constraining the equation of state of
  neutron stars from binary mergers},\ }\href
  {https://doi.org/10.1103/PhysRevLett.113.091104} {\bibfield  {journal}
  {\bibinfo  {journal} {Phys. Rev. Lett.}\ }\textbf {\bibinfo {volume} {113}},\
  \bibinfo {pages} {091104} (\bibinfo {year} {2014})}\BibitemShut {NoStop}%
\bibitem [{\citenamefont {Bauswein}\ and\ \citenamefont
  {Stergioulas}(2015)}]{PhysRevD.91.124056}%
  \BibitemOpen
  \bibfield  {author} {\bibinfo {author} {\bibfnamefont {A.}~\bibnamefont
  {Bauswein}}\ and\ \bibinfo {author} {\bibfnamefont {N.}~\bibnamefont
  {Stergioulas}},\ }\bibfield  {title} {\bibinfo {title} {Unified picture of
  the post-merger dynamics and gravitational wave emission in neutron star
  mergers},\ }\href {https://doi.org/10.1103/PhysRevD.91.124056} {\bibfield
  {journal} {\bibinfo  {journal} {Phys. Rev. D}\ }\textbf {\bibinfo {volume}
  {91}},\ \bibinfo {pages} {124056} (\bibinfo {year} {2015})}\BibitemShut
  {NoStop}%
\bibitem [{\citenamefont {Helstrom}(1967)}]{HELSTROM1967101}%
  \BibitemOpen
  \bibfield  {author} {\bibinfo {author} {\bibfnamefont {C.}~\bibnamefont
  {Helstrom}},\ }\bibfield  {title} {\bibinfo {title} {Minimum mean-squared
  error of estimates in quantum statistics},\ }\href
  {https://doi.org/https://doi.org/10.1016/0375-9601(67)90366-0} {\bibfield
  {journal} {\bibinfo  {journal} {Physics Letters A}\ }\textbf {\bibinfo
  {volume} {25}},\ \bibinfo {pages} {101 } (\bibinfo {year}
  {1967})}\BibitemShut {NoStop}%
\bibitem [{\citenamefont {Braginsky}\ \emph {et~al.}(2000)\citenamefont
  {Braginsky}, \citenamefont {Gorodetsky}, \citenamefont {Khalili},\ and\
  \citenamefont {Thorne}}]{Energetic}%
  \BibitemOpen
  \bibfield  {author} {\bibinfo {author} {\bibfnamefont {V.~B.}\ \bibnamefont
  {Braginsky}}, \bibinfo {author} {\bibfnamefont {M.~L.}\ \bibnamefont
  {Gorodetsky}}, \bibinfo {author} {\bibfnamefont {F.~Y.}\ \bibnamefont
  {Khalili}},\ and\ \bibinfo {author} {\bibfnamefont {K.~S.}\ \bibnamefont
  {Thorne}},\ }\bibfield  {title} {\bibinfo {title} {Energetic quantum limit in
  large-scale interferometers},\ }\href {https://doi.org/10.1063/1.1291855}
  {\bibfield  {journal} {\bibinfo  {journal} {AIP Conference Proceedings}\
  }\textbf {\bibinfo {volume} {523}},\ \bibinfo {pages} {180} (\bibinfo {year}
  {2000})}\BibitemShut {NoStop}%
\bibitem [{\citenamefont {Tsang}\ \emph {et~al.}(2011)\citenamefont {Tsang},
  \citenamefont {Wiseman},\ and\ \citenamefont
  {Caves}}]{PhysRevLett.106.090401}%
  \BibitemOpen
  \bibfield  {author} {\bibinfo {author} {\bibfnamefont {M.}~\bibnamefont
  {Tsang}}, \bibinfo {author} {\bibfnamefont {H.~M.}\ \bibnamefont {Wiseman}},\
  and\ \bibinfo {author} {\bibfnamefont {C.~M.}\ \bibnamefont {Caves}},\
  }\bibfield  {title} {\bibinfo {title} {Fundamental quantum limit to waveform
  estimation},\ }\href {https://doi.org/10.1103/PhysRevLett.106.090401}
  {\bibfield  {journal} {\bibinfo  {journal} {Phys. Rev. Lett.}\ }\textbf
  {\bibinfo {volume} {106}},\ \bibinfo {pages} {090401} (\bibinfo {year}
  {2011})}\BibitemShut {NoStop}%
\bibitem [{\citenamefont {Danilishin}\ \emph {et~al.}(2019)\citenamefont
  {Danilishin}, \citenamefont {Khalili},\ and\ \citenamefont
  {Miao}}]{Danilishin2019}%
  \BibitemOpen
  \bibfield  {author} {\bibinfo {author} {\bibfnamefont {S.~L.}\ \bibnamefont
  {Danilishin}}, \bibinfo {author} {\bibfnamefont {F.~Y.}\ \bibnamefont
  {Khalili}},\ and\ \bibinfo {author} {\bibfnamefont {H.}~\bibnamefont
  {Miao}},\ }\bibfield  {title} {\bibinfo {title} {Advanced quantum techniques
  for future gravitational-wave detectors},\ }\href
  {https://doi.org/10.1007/s41114-019-0018-y} {\bibfield  {journal} {\bibinfo
  {journal} {Living Reviews in Relativity}\ }\textbf {\bibinfo {volume} {22}},\
  \bibinfo {pages} {2} (\bibinfo {year} {2019})}\BibitemShut {NoStop}%
\bibitem [{\citenamefont {Miao}\ \emph {et~al.}(2019)\citenamefont {Miao},
  \citenamefont {Smith},\ and\ \citenamefont {Evans}}]{PhysRevX.9.011053}%
  \BibitemOpen
  \bibfield  {author} {\bibinfo {author} {\bibfnamefont {H.}~\bibnamefont
  {Miao}}, \bibinfo {author} {\bibfnamefont {N.~D.}\ \bibnamefont {Smith}},\
  and\ \bibinfo {author} {\bibfnamefont {M.}~\bibnamefont {Evans}},\ }\bibfield
   {title} {\bibinfo {title} {Quantum limit for laser interferometric
  gravitational-wave detectors from optical dissipation},\ }\href
  {https://doi.org/10.1103/PhysRevX.9.011053} {\bibfield  {journal} {\bibinfo
  {journal} {Phys. Rev. X}\ }\textbf {\bibinfo {volume} {9}},\ \bibinfo {pages}
  {011053} (\bibinfo {year} {2019})}\BibitemShut {NoStop}%
\bibitem [{\citenamefont {Ganapathy}\ \emph {et~al.}(2021)\citenamefont
  {Ganapathy}, \citenamefont {McCuller}, \citenamefont {Rollins}, \citenamefont
  {Hall}, \citenamefont {Barsotti},\ and\ \citenamefont
  {Evans}}]{PhysRevD.103.022002}%
  \BibitemOpen
  \bibfield  {author} {\bibinfo {author} {\bibfnamefont {D.}~\bibnamefont
  {Ganapathy}}, \bibinfo {author} {\bibfnamefont {L.}~\bibnamefont {McCuller}},
  \bibinfo {author} {\bibfnamefont {J.~G.}\ \bibnamefont {Rollins}}, \bibinfo
  {author} {\bibfnamefont {E.~D.}\ \bibnamefont {Hall}}, \bibinfo {author}
  {\bibfnamefont {L.}~\bibnamefont {Barsotti}},\ and\ \bibinfo {author}
  {\bibfnamefont {M.}~\bibnamefont {Evans}},\ }\bibfield  {title} {\bibinfo
  {title} {Tuning advanced ligo to kilohertz signals from neutron-star
  collisions},\ }\href {https://doi.org/10.1103/PhysRevD.103.022002} {\bibfield
   {journal} {\bibinfo  {journal} {Phys. Rev. D}\ }\textbf {\bibinfo {volume}
  {103}},\ \bibinfo {pages} {022002} (\bibinfo {year} {2021})}\BibitemShut
  {NoStop}%
\bibitem [{\citenamefont {Miao}\ \emph {et~al.}(2015)\citenamefont {Miao},
  \citenamefont {Ma}, \citenamefont {Zhao},\ and\ \citenamefont
  {Chen}}]{PhysRevLett.115.211104}%
  \BibitemOpen
  \bibfield  {author} {\bibinfo {author} {\bibfnamefont {H.}~\bibnamefont
  {Miao}}, \bibinfo {author} {\bibfnamefont {Y.}~\bibnamefont {Ma}}, \bibinfo
  {author} {\bibfnamefont {C.}~\bibnamefont {Zhao}},\ and\ \bibinfo {author}
  {\bibfnamefont {Y.}~\bibnamefont {Chen}},\ }\bibfield  {title} {\bibinfo
  {title} {Enhancing the bandwidth of gravitational-wave detectors with
  unstable optomechanical filters},\ }\href
  {https://doi.org/10.1103/PhysRevLett.115.211104} {\bibfield  {journal}
  {\bibinfo  {journal} {Phys. Rev. Lett.}\ }\textbf {\bibinfo {volume} {115}},\
  \bibinfo {pages} {211104} (\bibinfo {year} {2015})}\BibitemShut {NoStop}%
\bibitem [{\citenamefont {Page}\ \emph {et~al.}(2020)\citenamefont {Page},
  \citenamefont {Goryachev}, \citenamefont {Miao}, \citenamefont {Chen},
  \citenamefont {Ma}, \citenamefont {Mason}, \citenamefont {Rossi},
  \citenamefont {Blair}, \citenamefont {Ju}, \citenamefont {Blair},
  \citenamefont {Schliesser}, \citenamefont {Tobar},\ and\ \citenamefont
  {Zhao}}]{page2020gravitational}%
  \BibitemOpen
  \bibfield  {author} {\bibinfo {author} {\bibfnamefont {M.~A.}\ \bibnamefont
  {Page}}, \bibinfo {author} {\bibfnamefont {M.}~\bibnamefont {Goryachev}},
  \bibinfo {author} {\bibfnamefont {H.}~\bibnamefont {Miao}}, \bibinfo {author}
  {\bibfnamefont {Y.}~\bibnamefont {Chen}}, \bibinfo {author} {\bibfnamefont
  {Y.}~\bibnamefont {Ma}}, \bibinfo {author} {\bibfnamefont {D.}~\bibnamefont
  {Mason}}, \bibinfo {author} {\bibfnamefont {M.}~\bibnamefont {Rossi}},
  \bibinfo {author} {\bibfnamefont {C.~D.}\ \bibnamefont {Blair}}, \bibinfo
  {author} {\bibfnamefont {L.}~\bibnamefont {Ju}}, \bibinfo {author}
  {\bibfnamefont {D.~G.}\ \bibnamefont {Blair}}, \bibinfo {author}
  {\bibfnamefont {A.}~\bibnamefont {Schliesser}}, \bibinfo {author}
  {\bibfnamefont {M.~E.}\ \bibnamefont {Tobar}},\ and\ \bibinfo {author}
  {\bibfnamefont {C.}~\bibnamefont {Zhao}},\ }\href@noop {} {\bibinfo {title}
  {Gravitational wave detectors with broadband high frequency sensitivity}}
  (\bibinfo {year} {2020}),\ \Eprint {https://arxiv.org/abs/2007.08766}
  {arXiv:2007.08766 [physics.optics]} \BibitemShut {NoStop}%
\bibitem [{\citenamefont {Zhang}\ \emph
  {et~al.}(2021{\natexlab{b}})\citenamefont {Zhang}, \citenamefont {Bentley},\
  and\ \citenamefont {Miao}}]{galaxies9010003}%
  \BibitemOpen
  \bibfield  {author} {\bibinfo {author} {\bibfnamefont {T.}~\bibnamefont
  {Zhang}}, \bibinfo {author} {\bibfnamefont {J.}~\bibnamefont {Bentley}},\
  and\ \bibinfo {author} {\bibfnamefont {H.}~\bibnamefont {Miao}},\ }\bibfield
  {title} {\bibinfo {title} {A broadband signal recycling scheme for
  approaching the quantum limit from optical losses},\ }\bibfield  {journal}
  {\bibinfo  {journal} {Galaxies}\ }\textbf {\bibinfo {volume} {9}},\ \href
  {https://doi.org/10.3390/galaxies9010003} {10.3390/galaxies9010003} (\bibinfo
  {year} {2021}{\natexlab{b}})\BibitemShut {NoStop}%
\bibitem [{\citenamefont {Korobko}\ \emph {et~al.}(2019)\citenamefont
  {Korobko}, \citenamefont {Ma}, \citenamefont {Chen},\ and\ \citenamefont
  {Schnabel}}]{korobko2019quantum}%
  \BibitemOpen
  \bibfield  {author} {\bibinfo {author} {\bibfnamefont {M.}~\bibnamefont
  {Korobko}}, \bibinfo {author} {\bibfnamefont {Y.}~\bibnamefont {Ma}},
  \bibinfo {author} {\bibfnamefont {Y.}~\bibnamefont {Chen}},\ and\ \bibinfo
  {author} {\bibfnamefont {R.}~\bibnamefont {Schnabel}},\ }\bibfield  {title}
  {\bibinfo {title} {Quantum expander for gravitational-wave observatories},\
  }\href@noop {} {\bibfield  {journal} {\bibinfo  {journal} {Light: Science \&
  Applications}\ }\textbf {\bibinfo {volume} {8}},\ \bibinfo {pages} {1}
  (\bibinfo {year} {2019})}\BibitemShut {NoStop}%
\bibitem [{\citenamefont {Hild}\ and\ \citenamefont
  {Freise}(2007)}]{Hild_2007}%
  \BibitemOpen
  \bibfield  {author} {\bibinfo {author} {\bibfnamefont {S.}~\bibnamefont
  {Hild}}\ and\ \bibinfo {author} {\bibfnamefont {A.}~\bibnamefont {Freise}},\
  }\bibfield  {title} {\bibinfo {title} {A novel concept for increasing the
  peak sensitivity of {LIGO} by detuning the arm cavities},\ }\href
  {https://doi.org/10.1088/0264-9381/24/22/010} {\bibfield  {journal} {\bibinfo
   {journal} {Classical and Quantum Gravity}\ }\textbf {\bibinfo {volume}
  {24}},\ \bibinfo {pages} {5453} (\bibinfo {year} {2007})}\BibitemShut
  {NoStop}%
\bibitem [{\citenamefont {Huttner}\ \emph {et~al.}(2016)\citenamefont
  {Huttner}, \citenamefont {Danilishin}, \citenamefont {Barr}, \citenamefont
  {Bell}, \citenamefont {Gräf}, \citenamefont {Hennig}, \citenamefont {Hild},
  \citenamefont {Houston}, \citenamefont {Leavey}, \citenamefont {Pascucci},
  \citenamefont {Sorazu}, \citenamefont {Spencer}, \citenamefont
  {Steinlechner}, \citenamefont {Wright}, \citenamefont {Zhang},\ and\
  \citenamefont {Strain}}]{Huttner_2016}%
  \BibitemOpen
  \bibfield  {author} {\bibinfo {author} {\bibfnamefont {S.~H.}\ \bibnamefont
  {Huttner}}, \bibinfo {author} {\bibfnamefont {S.~L.}\ \bibnamefont
  {Danilishin}}, \bibinfo {author} {\bibfnamefont {B.~W.}\ \bibnamefont
  {Barr}}, \bibinfo {author} {\bibfnamefont {A.~S.}\ \bibnamefont {Bell}},
  \bibinfo {author} {\bibfnamefont {C.}~\bibnamefont {Gräf}}, \bibinfo
  {author} {\bibfnamefont {J.~S.}\ \bibnamefont {Hennig}}, \bibinfo {author}
  {\bibfnamefont {S.}~\bibnamefont {Hild}}, \bibinfo {author} {\bibfnamefont
  {E.~A.}\ \bibnamefont {Houston}}, \bibinfo {author} {\bibfnamefont {S.~S.}\
  \bibnamefont {Leavey}}, \bibinfo {author} {\bibfnamefont {D.}~\bibnamefont
  {Pascucci}}, \bibinfo {author} {\bibfnamefont {B.}~\bibnamefont {Sorazu}},
  \bibinfo {author} {\bibfnamefont {A.~P.}\ \bibnamefont {Spencer}}, \bibinfo
  {author} {\bibfnamefont {S.}~\bibnamefont {Steinlechner}}, \bibinfo {author}
  {\bibfnamefont {J.~L.}\ \bibnamefont {Wright}}, \bibinfo {author}
  {\bibfnamefont {T.}~\bibnamefont {Zhang}},\ and\ \bibinfo {author}
  {\bibfnamefont {K.~A.}\ \bibnamefont {Strain}},\ }\bibfield  {title}
  {\bibinfo {title} {Candidates for a possible third-generation gravitational
  wave detector: comparison of ring-sagnac and sloshing-sagnac speedmeter
  interferometers},\ }\href {https://doi.org/10.1088/1361-6382/34/2/024001}
  {\bibfield  {journal} {\bibinfo  {journal} {Classical and Quantum Gravity}\
  }\textbf {\bibinfo {volume} {34}},\ \bibinfo {pages} {024001} (\bibinfo
  {year} {2016})}\BibitemShut {NoStop}%
\bibitem [{\citenamefont {Huttner}\ \emph {et~al.}(2020)\citenamefont
  {Huttner}, \citenamefont {Danilishin}, \citenamefont {Hild},\ and\
  \citenamefont {Strain}}]{Huttner_2020}%
  \BibitemOpen
  \bibfield  {author} {\bibinfo {author} {\bibfnamefont {S.~H.}\ \bibnamefont
  {Huttner}}, \bibinfo {author} {\bibfnamefont {S.~L.}\ \bibnamefont
  {Danilishin}}, \bibinfo {author} {\bibfnamefont {S.}~\bibnamefont {Hild}},\
  and\ \bibinfo {author} {\bibfnamefont {K.~A.}\ \bibnamefont {Strain}},\
  }\bibfield  {title} {\bibinfo {title} {Comparison of different sloshing
  speedmeters},\ }\href {https://doi.org/10.1088/1361-6382/ab7bbb} {\bibfield
  {journal} {\bibinfo  {journal} {Classical and Quantum Gravity}\ }\textbf
  {\bibinfo {volume} {37}},\ \bibinfo {pages} {085022} (\bibinfo {year}
  {2020})}\BibitemShut {NoStop}%
\bibitem [{\citenamefont {Danilishin}\ and\ \citenamefont
  {Khalili}(2012)}]{SD2012}%
  \BibitemOpen
  \bibfield  {author} {\bibinfo {author} {\bibfnamefont {S.~L.}\ \bibnamefont
  {Danilishin}}\ and\ \bibinfo {author} {\bibfnamefont {F.~Y.}\ \bibnamefont
  {Khalili}},\ }\bibfield  {title} {\bibinfo {title} {Quantum measurement
  theory in gravitational-wave detectors},\ }\href
  {http://www.livingreviews.org/lrr-2012-5} {\bibfield  {journal} {\bibinfo
  {journal} {Living Reviews in Relativity}\ }\textbf {\bibinfo {volume} {15}}
  (\bibinfo {year} {2012})}\BibitemShut {NoStop}%
\bibitem [{\citenamefont {Purdue}\ and\ \citenamefont
  {Chen}(2002)}]{PhysRevD.66.122004}%
  \BibitemOpen
  \bibfield  {author} {\bibinfo {author} {\bibfnamefont {P.}~\bibnamefont
  {Purdue}}\ and\ \bibinfo {author} {\bibfnamefont {Y.}~\bibnamefont {Chen}},\
  }\bibfield  {title} {\bibinfo {title} {Practical speed meter designs for
  quantum nondemolition gravitational-wave interferometers},\ }\href
  {https://doi.org/10.1103/PhysRevD.66.122004} {\bibfield  {journal} {\bibinfo
  {journal} {Phys. Rev. D}\ }\textbf {\bibinfo {volume} {66}},\ \bibinfo
  {pages} {122004} (\bibinfo {year} {2002})}\BibitemShut {NoStop}%
\bibitem [{\citenamefont {Miao}\ \emph {et~al.}(2018)\citenamefont {Miao},
  \citenamefont {Yang},\ and\ \citenamefont {Martynov}}]{PhysRevD.98.044044}%
  \BibitemOpen
  \bibfield  {author} {\bibinfo {author} {\bibfnamefont {H.}~\bibnamefont
  {Miao}}, \bibinfo {author} {\bibfnamefont {H.}~\bibnamefont {Yang}},\ and\
  \bibinfo {author} {\bibfnamefont {D.}~\bibnamefont {Martynov}},\ }\bibfield
  {title} {\bibinfo {title} {Towards the design of gravitational-wave detectors
  for probing neutron-star physics},\ }\href
  {https://doi.org/10.1103/PhysRevD.98.044044} {\bibfield  {journal} {\bibinfo
  {journal} {Phys. Rev. D}\ }\textbf {\bibinfo {volume} {98}},\ \bibinfo
  {pages} {044044} (\bibinfo {year} {2018})}\BibitemShut {NoStop}%
\bibitem [{\citenamefont {{Danilishin}}\ \emph {et~al.}(2015)\citenamefont
  {{Danilishin}}, \citenamefont {{Gr{\"a}f}}, \citenamefont {{Leavey}},
  \citenamefont {{Hennig}}, \citenamefont {{Houston}}, \citenamefont
  {{Pascucci}}, \citenamefont {{Steinlechner}}, \citenamefont {{Wright}},\ and\
  \citenamefont {{Hild}}}]{2015asymSag}%
  \BibitemOpen
  \bibfield  {author} {\bibinfo {author} {\bibfnamefont {S.~L.}\ \bibnamefont
  {{Danilishin}}}, \bibinfo {author} {\bibfnamefont {C.}~\bibnamefont
  {{Gr{\"a}f}}}, \bibinfo {author} {\bibfnamefont {S.~S.}\ \bibnamefont
  {{Leavey}}}, \bibinfo {author} {\bibfnamefont {J.}~\bibnamefont {{Hennig}}},
  \bibinfo {author} {\bibfnamefont {E.~A.}\ \bibnamefont {{Houston}}}, \bibinfo
  {author} {\bibfnamefont {D.}~\bibnamefont {{Pascucci}}}, \bibinfo {author}
  {\bibfnamefont {S.}~\bibnamefont {{Steinlechner}}}, \bibinfo {author}
  {\bibfnamefont {J.}~\bibnamefont {{Wright}}},\ and\ \bibinfo {author}
  {\bibfnamefont {S.}~\bibnamefont {{Hild}}},\ }\bibfield  {title} {\bibinfo
  {title} {{Quantum noise of non-ideal Sagnac speed meter interferometer with
  asymmetries}},\ }\href {https://doi.org/10.1088/1367-2630/17/4/043031}
  {\bibfield  {journal} {\bibinfo  {journal} {New Journal of Physics}\ }\textbf
  {\bibinfo {volume} {17}},\ \bibinfo {eid} {043031} (\bibinfo {year}
  {2015})},\ \Eprint {https://arxiv.org/abs/1412.0931} {arXiv:1412.0931
  [quant-ph]} \BibitemShut {NoStop}%
\bibitem [{\citenamefont {Zhang}\ \emph {et~al.}(2018)\citenamefont {Zhang},
  \citenamefont {Knyazev}, \citenamefont {Steinlechner}, \citenamefont
  {Khalili}, \citenamefont {Barr}, \citenamefont {Bell}, \citenamefont {Dupej},
  \citenamefont {Briggs}, \citenamefont {Gräf}, \citenamefont {Callaghan},
  \citenamefont {Hennig}, \citenamefont {Houston}, \citenamefont {Huttner},
  \citenamefont {Leavey}, \citenamefont {Pascucci}, \citenamefont {Sorazu},
  \citenamefont {Spencer}, \citenamefont {Wright}, \citenamefont {Strain},
  \citenamefont {Hild},\ and\ \citenamefont {Danilishin}}]{Zhang_2018}%
  \BibitemOpen
  \bibfield  {author} {\bibinfo {author} {\bibfnamefont {T.}~\bibnamefont
  {Zhang}}, \bibinfo {author} {\bibfnamefont {E.}~\bibnamefont {Knyazev}},
  \bibinfo {author} {\bibfnamefont {S.}~\bibnamefont {Steinlechner}}, \bibinfo
  {author} {\bibfnamefont {F.~Y.}\ \bibnamefont {Khalili}}, \bibinfo {author}
  {\bibfnamefont {B.~W.}\ \bibnamefont {Barr}}, \bibinfo {author}
  {\bibfnamefont {A.~S.}\ \bibnamefont {Bell}}, \bibinfo {author}
  {\bibfnamefont {P.}~\bibnamefont {Dupej}}, \bibinfo {author} {\bibfnamefont
  {J.}~\bibnamefont {Briggs}}, \bibinfo {author} {\bibfnamefont
  {C.}~\bibnamefont {Gräf}}, \bibinfo {author} {\bibfnamefont
  {J.}~\bibnamefont {Callaghan}}, \bibinfo {author} {\bibfnamefont {J.~S.}\
  \bibnamefont {Hennig}}, \bibinfo {author} {\bibfnamefont {E.~A.}\
  \bibnamefont {Houston}}, \bibinfo {author} {\bibfnamefont {S.~H.}\
  \bibnamefont {Huttner}}, \bibinfo {author} {\bibfnamefont {S.~S.}\
  \bibnamefont {Leavey}}, \bibinfo {author} {\bibfnamefont {D.}~\bibnamefont
  {Pascucci}}, \bibinfo {author} {\bibfnamefont {B.}~\bibnamefont {Sorazu}},
  \bibinfo {author} {\bibfnamefont {A.}~\bibnamefont {Spencer}}, \bibinfo
  {author} {\bibfnamefont {J.}~\bibnamefont {Wright}}, \bibinfo {author}
  {\bibfnamefont {K.~A.}\ \bibnamefont {Strain}}, \bibinfo {author}
  {\bibfnamefont {S.}~\bibnamefont {Hild}},\ and\ \bibinfo {author}
  {\bibfnamefont {S.~L.}\ \bibnamefont {Danilishin}},\ }\bibfield  {title}
  {\bibinfo {title} {Quantum noise cancellation in asymmetric speed metres with
  balanced homodyne readout},\ }\href
  {https://doi.org/10.1088/1367-2630/aae86e} {\bibfield  {journal} {\bibinfo
  {journal} {New Journal of Physics}\ }\textbf {\bibinfo {volume} {20}},\
  \bibinfo {pages} {103040} (\bibinfo {year} {2018})}\BibitemShut {NoStop}%
\bibitem [{\citenamefont {Freise}\ \emph {et~al.}(2004)\citenamefont {Freise},
  \citenamefont {Heinzel}, \citenamefont {Lück}, \citenamefont {Schilling},
  \citenamefont {Willke},\ and\ \citenamefont {Danzmann}}]{Freise_2004}%
  \BibitemOpen
  \bibfield  {author} {\bibinfo {author} {\bibfnamefont {A.}~\bibnamefont
  {Freise}}, \bibinfo {author} {\bibfnamefont {G.}~\bibnamefont {Heinzel}},
  \bibinfo {author} {\bibfnamefont {H.}~\bibnamefont {Lück}}, \bibinfo
  {author} {\bibfnamefont {R.}~\bibnamefont {Schilling}}, \bibinfo {author}
  {\bibfnamefont {B.}~\bibnamefont {Willke}},\ and\ \bibinfo {author}
  {\bibfnamefont {K.}~\bibnamefont {Danzmann}},\ }\bibfield  {title} {\bibinfo
  {title} {Frequency-domain interferometer simulation with higher-order spatial
  modes},\ }\href {https://doi.org/10.1088/0264-9381/21/5/102} {\bibfield
  {journal} {\bibinfo  {journal} {Classical and Quantum Gravity}\ }\textbf
  {\bibinfo {volume} {21}},\ \bibinfo {pages} {S1067} (\bibinfo {year}
  {2004})}\BibitemShut {NoStop}%
\bibitem [{\citenamefont {Komori}\ \emph {et~al.}(2020)\citenamefont {Komori},
  \citenamefont {Ganapathy}, \citenamefont {Whittle}, \citenamefont {McCuller},
  \citenamefont {Barsotti}, \citenamefont {Mavalvala},\ and\ \citenamefont
  {Evans}}]{PhysRevD.102.102003}%
  \BibitemOpen
  \bibfield  {author} {\bibinfo {author} {\bibfnamefont {K.}~\bibnamefont
  {Komori}}, \bibinfo {author} {\bibfnamefont {D.}~\bibnamefont {Ganapathy}},
  \bibinfo {author} {\bibfnamefont {C.}~\bibnamefont {Whittle}}, \bibinfo
  {author} {\bibfnamefont {L.}~\bibnamefont {McCuller}}, \bibinfo {author}
  {\bibfnamefont {L.}~\bibnamefont {Barsotti}}, \bibinfo {author}
  {\bibfnamefont {N.}~\bibnamefont {Mavalvala}},\ and\ \bibinfo {author}
  {\bibfnamefont {M.}~\bibnamefont {Evans}},\ }\bibfield  {title} {\bibinfo
  {title} {Demonstration of an amplitude filter cavity at gravitational-wave
  frequencies},\ }\href {https://doi.org/10.1103/PhysRevD.102.102003}
  {\bibfield  {journal} {\bibinfo  {journal} {Phys. Rev. D}\ }\textbf {\bibinfo
  {volume} {102}},\ \bibinfo {pages} {102003} (\bibinfo {year}
  {2020})}\BibitemShut {NoStop}%
\bibitem [{\citenamefont {Barsotti}\ \emph {et~al.}(2018)\citenamefont
  {Barsotti}, \citenamefont {McCuller}, \citenamefont {Evans},\ and\
  \citenamefont {Fritschel}}]{T1800042}%
  \BibitemOpen
  \bibfield  {author} {\bibinfo {author} {\bibfnamefont {L.}~\bibnamefont
  {Barsotti}}, \bibinfo {author} {\bibfnamefont {L.}~\bibnamefont {McCuller}},
  \bibinfo {author} {\bibfnamefont {M.}~\bibnamefont {Evans}},\ and\ \bibinfo
  {author} {\bibfnamefont {P.}~\bibnamefont {Fritschel}},\ }\bibfield  {title}
  {\bibinfo {title} {The a+ design curve},\ }\href
  {https://dcc.ligo.org/LIGO-T1800042} {\bibfield  {journal} {\bibinfo
  {journal} {LIGO Document T1800042}\ } (\bibinfo {year} {2018})}\BibitemShut
  {NoStop}%
\bibitem [{\citenamefont {Sanders}\ and\ \citenamefont
  {Ballmer}(2016)}]{Sanders_2016}%
  \BibitemOpen
  \bibfield  {author} {\bibinfo {author} {\bibfnamefont {J.~R.}\ \bibnamefont
  {Sanders}}\ and\ \bibinfo {author} {\bibfnamefont {S.~W.}\ \bibnamefont
  {Ballmer}},\ }\bibfield  {title} {\bibinfo {title} {Folding
  gravitational-wave interferometers},\ }\href
  {https://doi.org/10.1088/1361-6382/34/2/025003} {\bibfield  {journal}
  {\bibinfo  {journal} {Classical and Quantum Gravity}\ }\textbf {\bibinfo
  {volume} {34}},\ \bibinfo {pages} {025003} (\bibinfo {year}
  {2016})}\BibitemShut {NoStop}%
\end{thebibliography}%
\end{document}